\documentclass[final]{solarphysics}
\usepackage[optionalrh,solaenum,natbib]{spr-sola-addons} 
\usepackage{graphicx}                    
\usepackage{color}                       
\usepackage{url}                         
\usepackage{ textcomp }

\begin{document}

\begin{article}

\begin{opening}

\title{Observations of Low Frequency Solar Radio Bursts from the Rosse Solar-Terrestrial Observatory}

%
\author{P.~\surname{Zucca}$^{1}$\sep
        E.~P.~\surname{Carley}$^{1}$\sep
        J.~\surname{McCauley}$^{1}$\sep
        P. ~T.~\surname{Gallagher}$^{1}$\sep
        C.~\surname{Monstein}$^{2}$\sep
        R.~T.~J.~\surname{McAteer}$^{1,3}$      
       }
%

%
  \institute{$^{1}$ School of Physics, Trinity College Dublin, Dublin 2, Ireland.
                   email: \url{zuccap@tcd.ie} \\
                  $^{2}$Institute for Astronomy, ETH-Zentrum, Z\"{u}rich, CH-8093, Switzerland.\\
                  $^{3}$ Department of Astronomy, New Mexico State University, PO Box 30001 MSC 4500, Las Cruces, NM 88003, USA.\\ 
             }

\begin{abstract}
The Rosse Solar-Terrestrial Observatory (RSTO;  www.rosseobservatory.ie) was established at Birr Castle, Co. Offaly, Ireland (53\textdegree 05\textquotesingle38.9\textquotesingle \textquotesingle, 7\textdegree 55\textquotesingle 12.7\textquotesingle \textquotesingle) in 2010 to study solar radio bursts and the response of the Earth's ionosphere and geomagnetic field. 
To date, three \textit{Compound Astronomical Low-cost Low-frequency Instrument for Spectroscopy and Transportable Observatory} (CALLISTO) spectrometers have been installed, with the capability of observing in the frequency range 10--870~MHz. The receivers are fed simultaneously by biconical and log-periodic antennas. Nominally, frequency spectra in the range 10--400 MHz are obtained with 4 sweeps per second over 600 channels. Here, we describe the RSTO solar radio spectrometer set-up, and present dynamic spectra of a sample of Type II, III and IV radio bursts. In particular, we describe fine-scale structure observed in Type II bursts, including band splitting and rapidly varying herringbone features. 
\end{abstract}

%

\end{opening}

%
 \section{Introduction}
 
The Sun is an active star that produces large-scale energetic events, such as solar flares and coronal mass ejections (CMEs). These phenomena are observable across the electromagnetic spectrum, from gamma rays at hundreds of MeV to radio waves with wavelengths of tens of metres. Solar flares and CMEs can excite plasma oscillations which can emit radiation at metric and decametric wavelengths. These bursts are classified in five main types. Type I bursts are short-duration narrowband features that are associated with active regions \citep{Melrose1975}. Type II bursts are thought to be excited by magnetohydrodynamic (MHD) shockwaves associated with CMEs \citep{Nelson1985}, while Type IIIs result from energetic particles escaping along open magnetic field lines. Type IV bursts show broad continuum emission with rapidly-varying fine structures \citep{Labrum1985}. The smooth short-lived continuum following a Type III burst is called a Type V. 

Several radio telescope designs have been developed to observe solar radio activity, including interferometers, spectrometers and imaging-spectrometers. The \textit{Culgoora Radioheliograph} was a 96-element, 3~km diameter radio synthesis telescope and operated at 80 MHz \citep{Culgora1972}. Operations began in 1968, but were discontinued in 1986. However, Culgoora still operates a radio spectrograph at 18--1800~MHz. Developed in the early 1980s, the \textit{Nan\c{c}ay Decameter Array} consists of two phased arrays producing dynamic spectra at 10--80~MHz  \citep{Dumas1982}. This operates alongside a radioheliograph, which produces solar images at a number of frequencies between 150--420~MHz. During the 1980s and 1990s, ETH Zurich developed a number of solar radio spectrometers  \citep{Benz1991,Messmer1999}. Their latest instrument, \textit{Phoenix II}, is a Fourier-based spectrograph operating at 0.1--4~GHz, with 2000 channels and a temporal resolution of better than 1~s. In recent years, the \textit{Green Bank Solar Radio Burst Spectrometer} (GBRBS) has been developed in the US, composed of three swept-frequency systems that support observations at 18--1100~MHz, with a temporal resolution of approximately 1~ms \citep{White2007}.   Similarly, the \textit{Artemis} spectrograph in Greece, observes at 20 to 650~MHz, and operates with a 7~m moving parabolic antenna at 110--650~MHz and a stationary antenna for the 20--110~MHz range \citep{Kontogeorgos2006}. A new milestone in low frequency radio instrumentation was reached with the establishment of the Dutch-lead LOw Frequency ARray \citep[LOFAR;][]{deVos2009}. This is the latest development in software-based radio interferometry, and provides the capability of simultaneously recording dynamic spectra and images of solar phenomena \citep{Fallows2012}.

The CALLISTO \textit{(Compound Astronomical Low-cost Low-frequency Instrument for Spectroscopy and Transportable Observatory)} spectrograph is a new concept for solar radio spectrographs, designed by ETH Zurich \citep{Benz2005}. This is a low-cost radio spectrometer used to monitor metric and decametric radio bursts, and which has been deployed to a number of sites around the world to allow for 24 hour monitoring of solar radio activity. In order to monitor solar activity and its effects on the Earth, we set up an autonomous solar radio observing station, the Rosse Solar-Terrestrial Observatory (RSTO), which has been operating since September 2010. RSTO is located in the grounds of Birr Castle, Co. Offaly, Ireland, and was named for the 3rd Earl of Rosse, Sir William Parsons, who constructed the 6-feet diameter \textit{``Leviathan Telescope"} in the 1840s \citep{Leviathan2002}. 

\begin{figure} 
 \centerline{\includegraphics[width=1.2\textwidth,clip=]{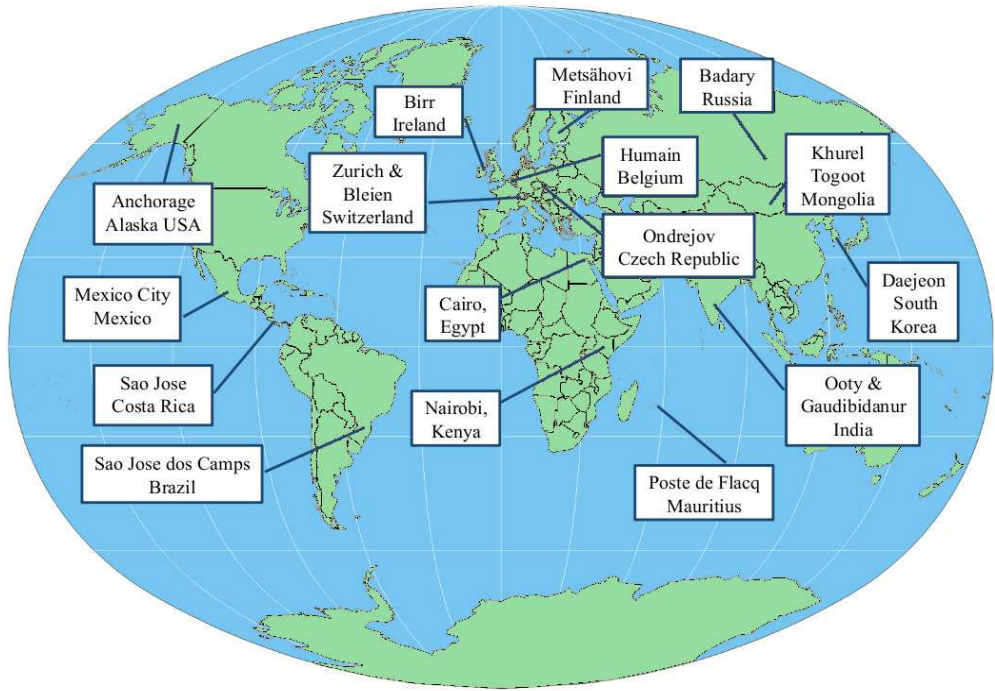}}
 \caption{Worldwide distribution of a portion of the radio spectrographs in the e-CALLISTO network. The network also includes spectrographs in Australia (Perth and Melbourne), Hawaii, Germany, Kazakhstan, Sri Lanka, Italy, Slovakia and Malaysia.}
 \label{worldwide}
 \end{figure}

RSTO is part of the e-CALLISTO network\footnote{www.e-CALLISTO.org}. The network consists of a number of spectrometers located around the globe, and designed to monitor solar radio emission in the metre and decametre bands (\citealt{Benz2009}; Figure~\ref{worldwide}).  Each of the instruments observes automatically, and data is collected each day via the Internet and stored in a central database at Fachhochschule Nordwestschweiz (FHNW), and operated by ETH Zurich\footnote{soleil.i4ds.ch/solarradio/CALLISTOQuicklooks/}. One of the important features of RSTO is the particularly low radio frequency interference (RFI) of the site, which is further described in Section~\ref{survey}.

In this paper, we describe the suite of CALLISTO spectrographs at RSTO (Section 2). Examples of observations are then reported in Section 3, while plans for future instruments at RSTO and conclusions are given in Section 4.
\newpage

\section{Radio Spectrometer Instrumentation}

\subsection{CALLISTO Spectrometer}

\begin{figure} 
 \centerline{\includegraphics[trim=1cm 1cm 6cm 1.5cm, clip=true,angle=-90,width=1.2\textwidth,clip=]{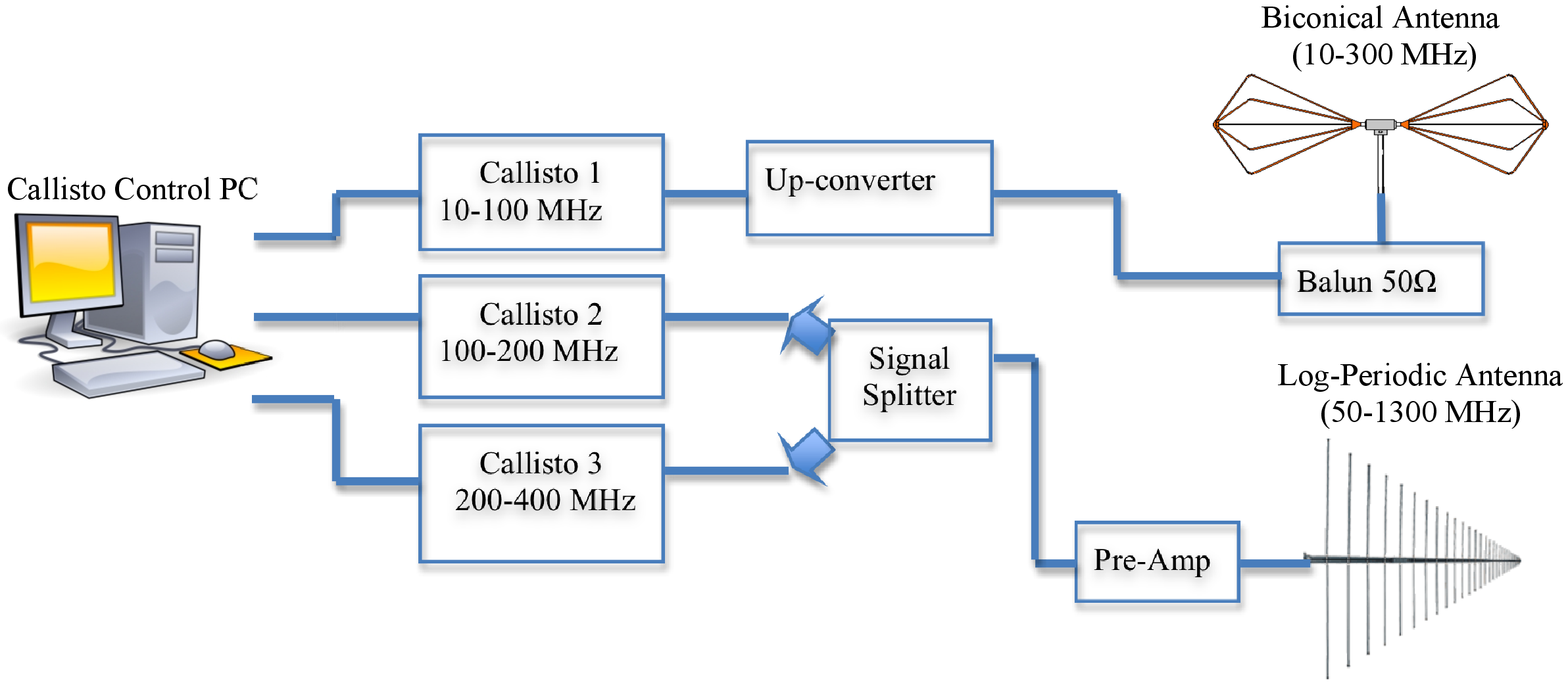}}
 \caption{The set-up of the array of three CALLISTO spectrographs  at RSTO. The current set-up employes three CALLISTO receivers, one connected to a bicone antenna using a frequency up-converter and measuring from 10 to 100~MHz. The other two receivers are connected to a log-periodic antenna measuring from 100 to 200~MHz and 200 to 400~MHz. The system can potentially observe between 10 and 870~MHz. }%
 \label{rstoset-up}
 \end{figure}
 
CALLISTO spectrometers are designed to monitor solar radio bursts in the frequency range 10--870~MHz. CALLISTO is composed of standard electronic components, employing a  Digital Video Broadcasting-Terrestrial (DVB-T) tuner assembled on a single printed circuit board. The number of channels per frequency sweep can vary between 1 and 400, with a maximum of 800 measurements per second. An individual channel has 300~kHz bandwidth during a typical frequency sweep of 250~ms, and can be tuned by the control software in steps of 62.5~kHz to obtain a more detailed spectrum of the radio environment. The narrow channel width allows for the measurement of selected channels that avoid known bands of radio interference from terrestrial sources. 

\subsection{CALLISTO Spectrometer Set-up at RSTO}

RSTO operates three CALLISTO receivers fed by a broadband log-periodic antenna and a biconical antenna (Figure~\ref{rstoset-up}). Nominally, the RSTO set-up operates at 600 channels with a sampling time of 250~ms seconds per sweep. CALLISTO 1 observes at 10--100~MHz, CALLISTO 2 at 100--200~MHz, and CALLISTO 3 at 200--400~MHz. The system has been optimised to measure the dynamic spectra of Type II radio bursts produced by coronal shock waves, and Type III radio bursts produced by near-relativistic electrons streaming along open magnetic field lines. It can also record other radio bursts, such as Type IV bursts and Type I noise storms. 

The log-periodic antenna has a frequency band of 50 to 1300~MHz with a $\sim$50 degree half-power beam-width (HPBW). The antenna is fixed to an alt-azimuth drive which tracks the Sun to optimize its response. The biconical antenna is 4~m long and has a nominal frequency sensitivity from 10 to 300~MHz. It is also mounted on a motorized rotator to track the Sun. CALLISTO 2 and 3 operates with a pre-amplifier that has a frequency range of 5--1500~MHz, and a typical noise figure of 1.2~dB, while a similar pre-amplifier is separately connected to CALLISTO 1. The system set-up is optimized to reach the ionospheric cutoff frequency at  $\sim$10~MHz. In order to do this, the receiver with a nominal operational band between 45 to 870~MHz has to operate with a frequency up-converter, shifting the range between 10--100~MHz to 220--310~MHz. The observed frequencies are then down-converted in software.

\begin{figure} 
 \centerline{\includegraphics[trim=0cm 0cm 0cm 0cm,angle=0,width=1\textwidth,clip=]{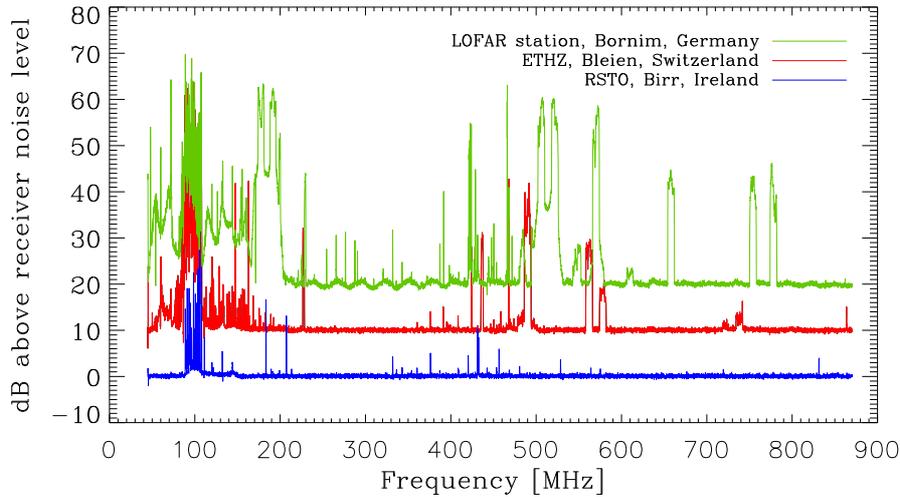}}
 \caption{Radio frequency survey of the RSTO in Birr Castle Demesne (blue), Bleien Radio Observatory in Switzerland (red; offset by 10 dB) and from Potsdam Bornim (green; offset by 20 dB). The RSTO spectrum is quiet at all frequencies that were tested, except for the FM band covering 88--108~MHz. The surveys were conducted using the same equipment. Note, LOFAR operates at $\sim$20--240~MHz.}
 \label{radiosurvey}
 \end{figure}

\newpage

\section{Observations}
\subsection{RSTO Radio Frequency Interference Survey}
\label{survey} 
A survey of RFI at RSTO was performed in June 2009\footnote{www.rosseobservatory.ie/presentations/birr\textunderscore radio\textunderscore survey.pdf}. The detected spectrum is shown in Figure \ref{radiosurvey}.   A commercial DVB-T antenna covering the range from 20~MHz up to 900~MHz was used for the survey, which was directly connected via a low-loss coaxial cable to a CALLISTO receiver with a sensitivity of 25~mV/dB. The channel resolution was 62.5~kHz, while the radiometric bandwidth was about 300 kHz. The sampling time was 1.25 ms per frequency interval, while the integration time was about 1~ms.  Figure \ref{radiosurvey} shows the RFI radio surveys of RSTO,  Bleien Observatory in Switzerland, and the Potsdam LOFAR station in Germany. There is an high level of interference at 20--200~MHz for the Bleien and Potsdam sites, while the RSTO site has a low level of RFI.

The radio spectrum at RSTO is extremely quiet compared to the majority of e-CALLISTO sites around the world. FM-radio and DVB-T are less intense than other sites, making the RSTO site an ideal location for low--frequency solar radio observations. Indeed, Birr Castle Demesne is a near--ideal site for frequency-agile or  Fourier-based spectrometers. All protected frequencies for radio astronomy are free from interference, and could be used for single frequency observations to determine solar radio flux using broadband antennas. As a result of this survey, the Irish astronomy community are considering Birr Castle Demesne as a site for a LOFAR station\footnote{www.lofar.ie}.

 \begin{figure} 
 \centerline{\includegraphics[trim=0cm 0cm 0cm 0cm,angle=0,width=1\textwidth,clip=]{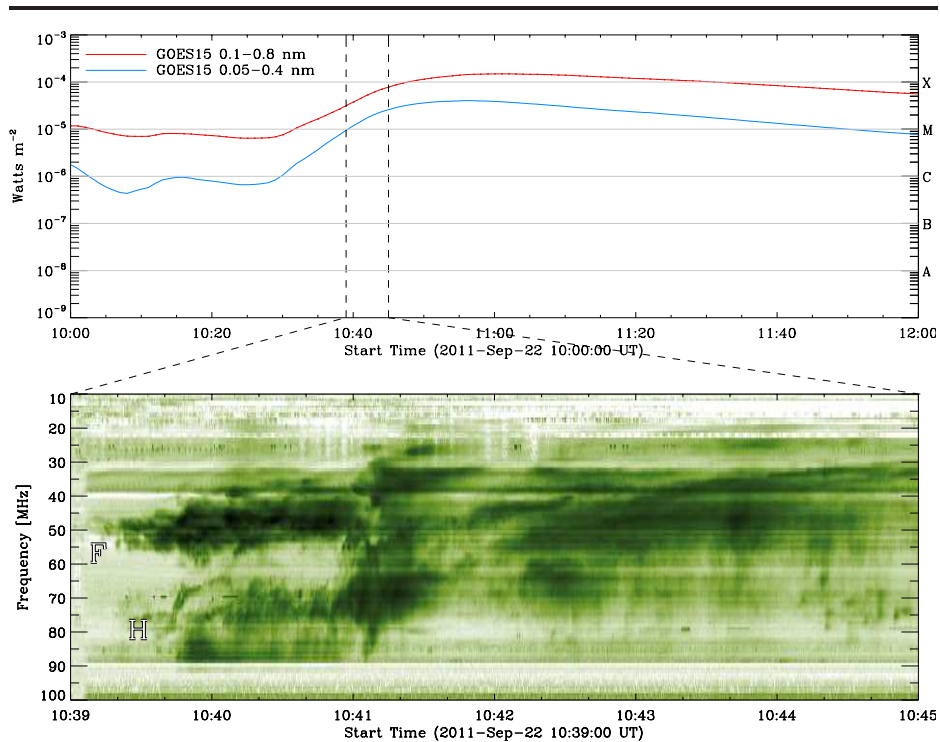}}
 \caption{Dynamic spectrum of the 22 September 2011 Type II radio burst and related GOES--15 light curve showing an X1.4 flare. This burst shows fundamental (F) and harmonic (H) emission. Band splitting of the order of 10~MHz can also be seen in the harmonic backbone at times around 10:42~UT.}
 \label{typeIIsplit}
 \end{figure}

\subsection{Sample RSTO Dynamic Spectra}
Observations started in September 2010, and first light was achieved on 17 November 2010. Since then, a large number of radio bursts have been recorded. In this section, we present a number of observations and give a brief description of each. All RSTO data is provided to the community at www.rosseobservatory.ie.
\begin{figure} 
 \centerline{\includegraphics[trim=0cm 0cm 0cm 0cm,angle=0,width=1\textwidth,clip=]{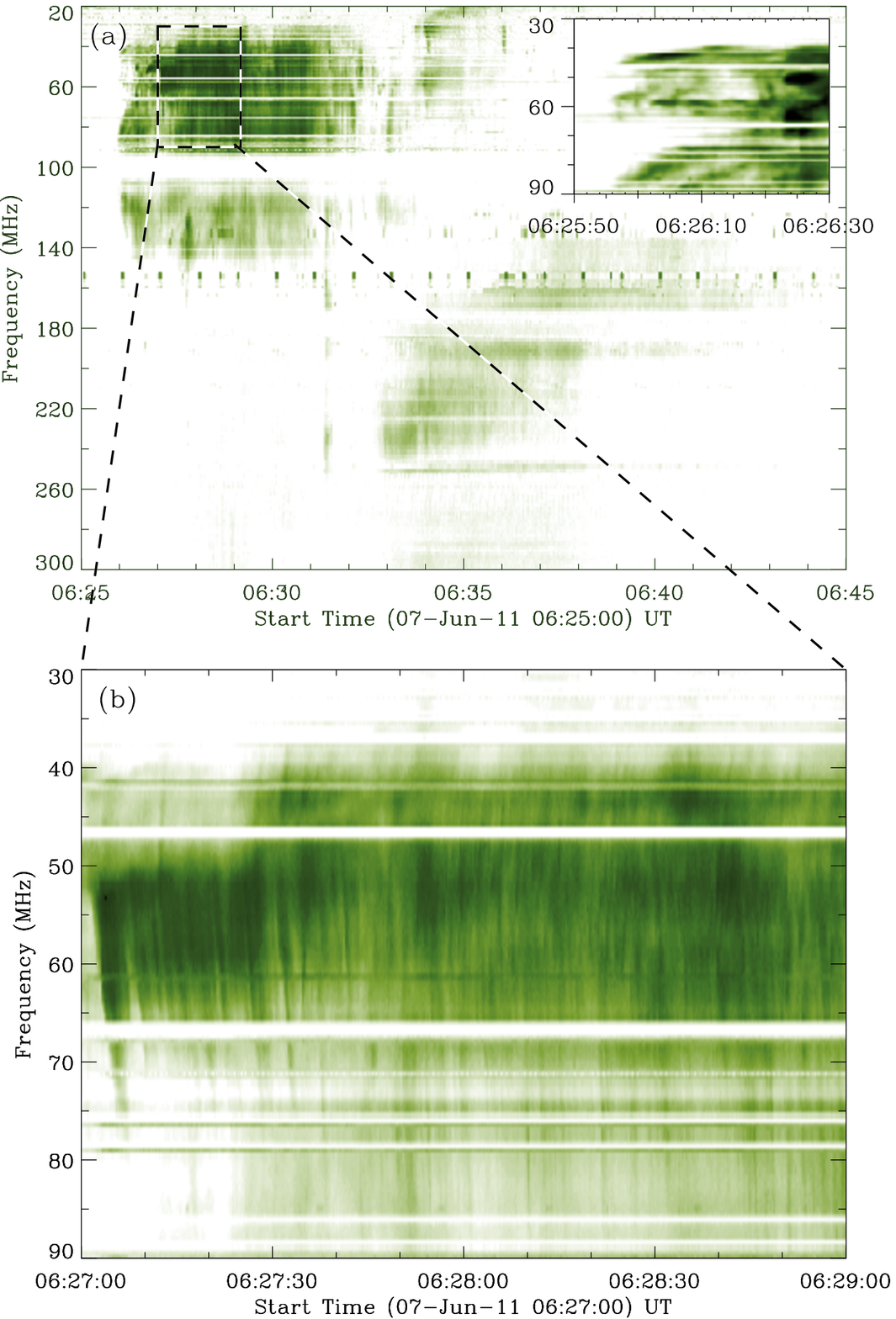}}
 \caption{Dynamic spectrum of the 07June 2011 Type II radio burst. In the inset of panel (a) a short-lived fundamental and harmonic backbones are visible. Multiple herringbone structure are visible at $\sim$40--80~MHz in panel (b).}
 \label{shorttypeII}
 \end{figure}
  
\subsubsection{Type II bursts}
The appearance of Type II radio bursts can vary significantly in dynamic spectra. The 22 September 2011 Type II radio burst shown in Figure~\ref{typeIIsplit} was associated with an X1.4 class flare which started at 10:29:00~UT. The flare was identified to have originated in NOAA active region 11302 and was associated with a CME. The burst started at 10:39:06 UT, and shows both fundamental (F) and harmonic (H) bands of emission. The fundamental emission is visible between 20 and 60~MHz, while the harmonic backbone lies between 60 and 90~MHz. Emission higher then 88~MHz is attenuated by the presence of the FM band. This structure is typical of the majority of Type II burst, i.e., when the harmonic backbone is present, it is almost always stronger than the fundamental. The two backbones show a drift rate of $\sim$0.22 MHz~s$^{-1}$, drifting towards lower frequencies as the plasma becomes less dense at larger distances from the Sun. A shock velocity of 1240 km~s$^{-1}$ was estimated using the 1--fold Newkirk model \citep{Newkirk1961}. 

Type II bursts typically last 5--10 minutes, but bursts exceeding 30 minutes have been known. Furthermore, short--duration bursts under one minute have been identified. A short--lived Type II observed on 2011 June 7 is shown in the inset on the top right of Figure~\ref{shorttypeII}(a). It is much more difficult to interpret short--duration Type II bursts, particularly if they occur at similar times and frequencies as other radio activity. In addition, the fundamental/harmonic structure is split into two roughly parallel bands as evident in Figure~\ref{shorttypeII}(a). The band-splitting phenomenon has two interpretations. It could be due to either the shape of the electron density distribution in the corona \citep{McLean1967} or the emission ahead and behind the shock  front \citep{Smerd1974}. 

\begin{figure} 
 \centerline{\includegraphics[angle=-0,width=1\textwidth,clip=]{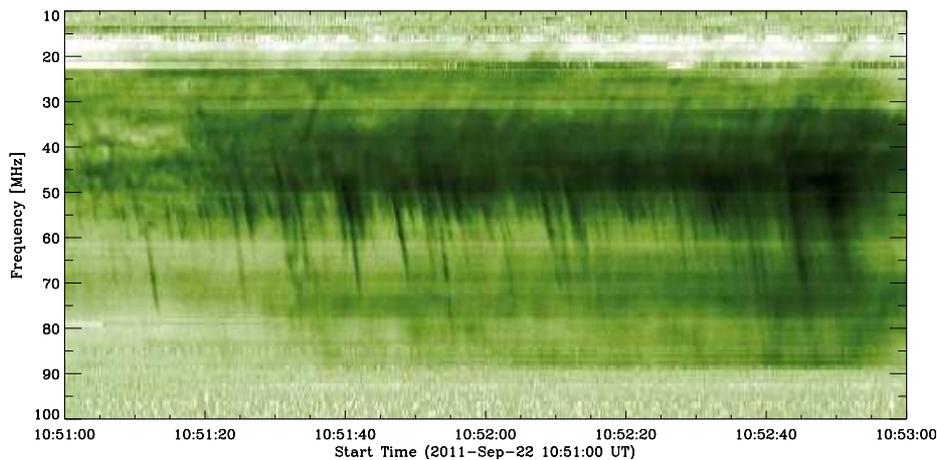}}
 \caption{Dynamic spectrum of herringbones following a Type II radio burst on 22 September 2011. These are thought to be due to electron beams shooting out from the shock front moving through the upper and lower corona.}
 \label{herringbones}
 \end{figure}

Another peculiarity of Type II burst is the sporadic presence of herringbones, small features similar to Type III bursts that straddle the backbones emission. In Figure~\ref{shorttypeII}(b) and Figure~\ref{herringbones}, herringbone features are evident. On 22 September 2011, about 50 herringbones drifting upwards in frequency between 40 and 80 MHz and downwards between 40 and 15 MHz are clearly visible between 10:51:00 UT and 10:53:00 UT. These are believed to be due to electron beams ejected out from the shock front moving through the upper and lower corona \citep{Holman1983}. The drift in frequency is very fast ($\sim$22 MHz~s$^{-1}$), with a corresponding velocity of $\sim$0.1c. 

 \begin{figure} 
 \centerline{\includegraphics[angle=0,width=1.2\textwidth,clip=]{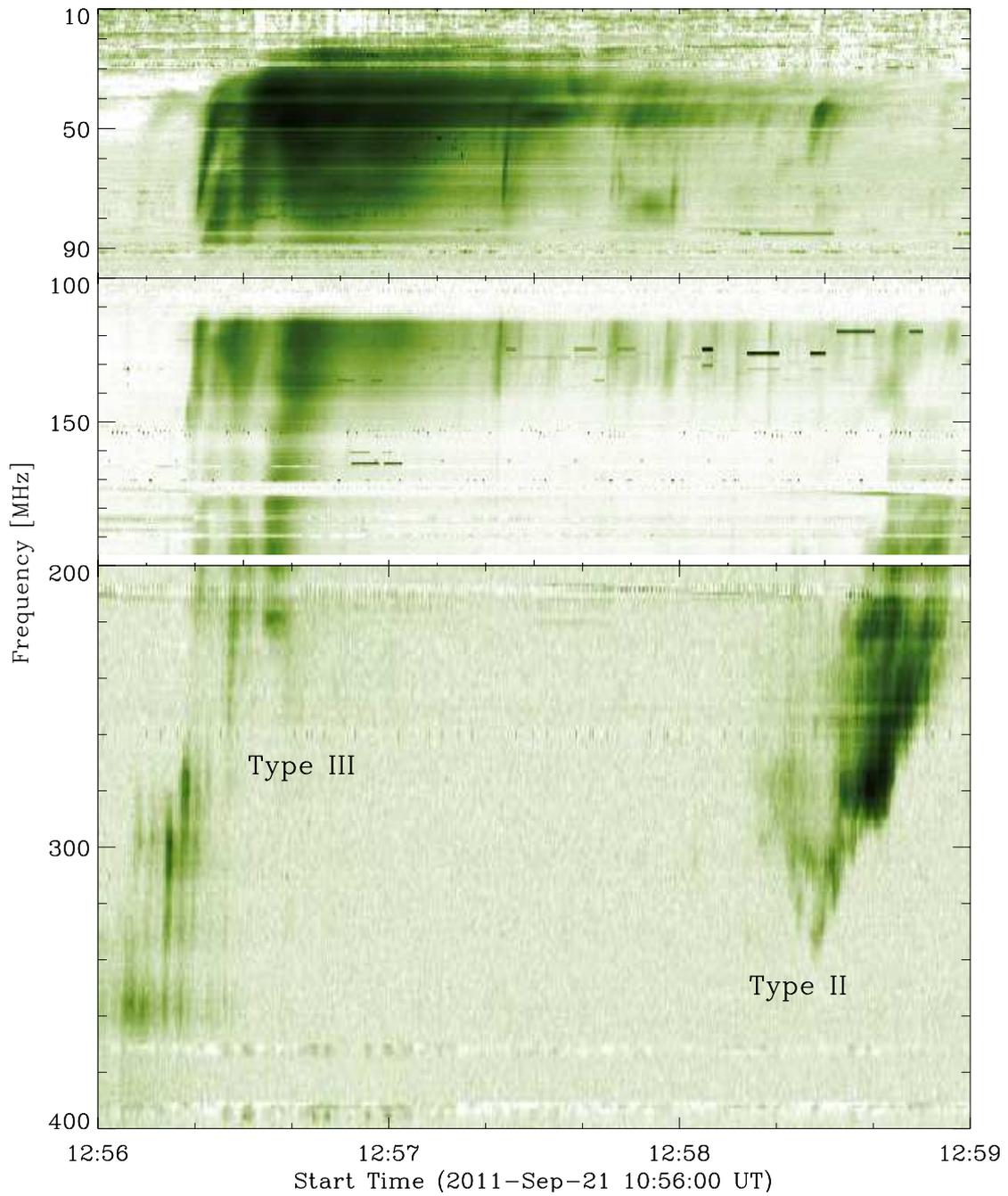}}
 \caption{Several Type III radio bursts observed on 21 October 2011. Broad band emission is superimposed on the bursts. Also shown a Type II burst between 140 and 330~MHz.}
 \label{TypeIII1}
 \end{figure}

\subsubsection{Type III bursts}
Figure~\ref{TypeIII1} shows a series of Type III bursts starting at 12:56:05 UT on 21 October 2011. The emission drifts from 400 to 20 MHz in frequency. A broad-band emission following the Type IIIs, called a Type V radio burst, is also evident from 20 to 150 MHz.  The event was associated with an M1 flare which occurred in NOAA 11319. Furthermore, an associated Type II burst starts at 12:58:12 UT, drifting from frequencies of 350 MHz to 170~MHz over 45~s. We determined a velocity drift using the 1--fold Newkirk Model of 790 km~s$^{-1}$. This Type II burst was possibly associated with a CME that appeared in the  SOHO/LASCO C2 field--of--view at 13:36:00 UT.  

Type III bursts can occur in groups, recurrently over an extended period, and continuously in the form of storms. Type III bursts are very common features in the metric range \citep{Cane2003}. Since the first-light of the CALLISTO-based spectrometer at RSTO, a large number of Type III bursts were detected. Type III bursts are produced by relativistic electrons traveling along open magnetic fields and therefore the drift in frequency detected in the radio spectra is very steep as the electrons travel fast in the corona and density becomes more and more tenuous. Similarly to Type II bursts, Type IIIs can often show a harmonic component. Since the drift rate is very fast, the two components usually merge, making their detection difficult \citep{Labrum1985}. 

 \begin{figure} 
 \centerline{\includegraphics[angle=0,width=1\textwidth,clip=]{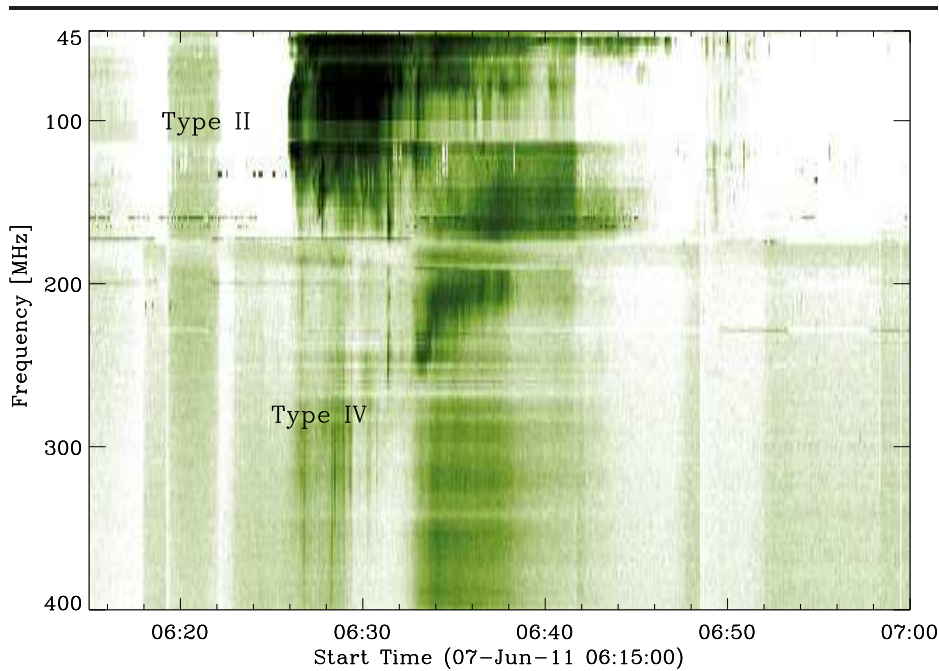}}
 \caption{Type II and Type IV radio bursts observed on 07June 2011. The darker feature starting at 06:25:30 between 150 and 45 MHz is a Type II radio burst. The spectrum shows a continuum emission starting at 06:31:10 UT between 400-130 MHz and a moving Type IV starting at 290 MHz drifting downwards in frequency. }
 \label{TypeIV}
 \end{figure}
 
\subsubsection{Type IV bursts}
In Figure~\ref{TypeIV}, a Type IV burst observed on 07June 2011 is shown. The emission is related to an M2 flare which occurred in NOAA 11226. A halo CME with an estimated plane-of-sky speed of 1155 km~s$^{-1}$ was also detected by LASCO. The broad emission started at 06:32:10 UT from 400 to 200 MHz and then spread from 400 to 100 MHz as it gained intensity. The burst can be seen to extend to 400 MHz for its duration, which is the upper frequency limit of the spectrograph. A drifting feature is also evident. This starts at about the same time as the Type IV, but which moves to lower frequencies over time. This is called a moving Type IV and has a bandwidth of approximately 100~MHz.  

Type IV continua can exhibit a wide range of forms \citep{Takakura1961}. They are broadband, usually 500 MHz wide, but some can exceed two or three times this. Type IV bursts can be very uniform in intensity, or they can fluctuate with complex underlying internal structures. Small fine structure are superimposed in the drifting continua. They are believed to be generated by emission of electrons trapped in post-flare loops and the drift is linked with the formation of the loops at successively higher altitudes \citep{Dulk1971}. 

\section{Discussion and Future Directions}
Three CALLISTO spectrographs are currently installed at RSTO, sampling the radio spectrum between 10 and 400 MHz with 600 frequency channels and a temporal resolution of 250~ms per sweep. The system can potentially measure between 10 and 870 MHz. This unique configuration of CALLISTO receivers and the use of a biconical antenna  and a frequency up-converter allows a good sampling of the spectra with a low-cost system. Since September 2010, a number of radio bursts have been detected, enabling us to identify fine-scale features in Type II and Type III radio bursts, including Type II band splitting and herringbones. One important characteristic of the site is its extremely low RFI. In 2012, a fourth CALLISTO receiver will be added to RSTO set-up in order to extend the current observational mode to 10--870~MHz, and increasing the number of channels to 800 per sweep.

We have also installed a ionispheric monitoring instrument as a complimentary instrument to CALLISTO at RSTO. The \textit{Atmospheric Weather Electromagnetic System of Observation, Modeling, and Education} \citep[AWESOME;][]{Cohen2010}. AWESOME is a ionospheric monitoring sensor designed by Stanford University, which is used to detect ionospheric disturbances. The Very Low Frequency (VLF) and Extremely Low Frequency (ELF) sensors of AWESOME monitor radio frequencies in the region of 3-30 kHz and 0.3-3 kHz, respectively. The AWESOME receiver uses wire-loop antennas, each sensitive to the component of the magnetic field in the direction orthogonal to the plane of the loop. ELF/VLF remote sensing enables study of a broad set of phenomena, each of which impacts the ionosphere in a unique way, including solar flares, cosmic gamma ray bursts, lightning strikes and lightning-related effects, earthquakes, electron precipitation, and the aurora. Sudden ionospheric disturbances (SIDs) occur in association with solar flares and have a very strong and relatively long-lasting effect on the ionosphere \citep{Cliverd2001}. 

We are now in the process of testing a flux-gate magnetometer to measure fluctuations in the Earth's magnetic field in Ireland. Using e-CALLISTO, we will measure the on-set time of solar radio bursts near the Sun, measure their effects on the Earth's ionosphere with AWESOME, and ultimately determine their effects to the geomagnetic field in Ireland. As part of a worldwide network of observatories, RSTO will provide an extensive capability to monitor solar activity and its effects on Earth in a real-time, continuous manner. This is made possible by the low RFI of the Birr Castle Demesne site. RSTO has been earmarked as an ideal location for a LOFAR station (www.lofar.ie). 
When combined with data from space-based observatories, such as NASA's STEREO and \textit{Solar Dynamic Observatory} (SDO) spacecraft, this will contribute towards a capability  to track and understand the propagation of storms from the surface of the Sun to their local effects on Earth.

\begin{acks}

The authors are indebted to the 7th Earl of Rosse and the Birr Castle Demesne staff, particularly George Vaugh, for their support during the development of the RSTO.  We would also like to thank the TCD Centre for Telecommunications Value-chain Research  (CTVR) and the School of Physics Mechanical Workshop. PZ is supported by a TCD Innovation Bursary. EC is a Government of Ireland Scholar supported by the Irish Research Council for Science, Engineering and Technology. We would also like to thank the Alice Barklie Bequest to the TCD School of Physics. RSTO was established under the auspices of International Heliophysical Year 2007 and the International Space Weather Initiative, supported by the United Nations Basic Space Science Initiative.
\end{acks}



\bibliographystyle{spr-mp-sola}
\bibliography{solar_physics_bib}  
\end{article} 
\end{document}